\documentclass[preprint]{aastex}
\usepackage{amsmath, amsthm}
\usepackage{graphics}
\usepackage{epsfig}
\usepackage{graphicx}
\usepackage{rotate}
\usepackage{color}

\shorttitle{Explosions of massive stars with 
magnetic winds}
\shortauthors{Biermann et al.}
\begin{document}
\baselineskip=24pt

\newcommand{\D}{\displaystyle} 
\newcommand{\T}{\textstyle} 
\newcommand{\SC}{\scriptstyle} 
\newcommand{\SSC}{\scriptscriptstyle} 

\def\AJ{{\it Astron. J.} }
\def\ARAA{{\it Annual Rev. of Astron. \& Astrophys.} }
\def\ApJ{{\it Astrophys. J.} }
\def\ApJL{{\it Astrophys. J. Letters} }
\def\ApJS{{\it Astrophys. J. Suppl.} }
\def\ApP{{\it Astropart. Phys.} }
\def\AA{{\it Astron. \& Astroph.} }
\def\AAR{{\it Astron. \& Astroph. Rev.} }
\def\AAL{{\it Astron. \& Astroph. Letters} }
\def\AASu{{\it Astron. \& Astroph. Suppl.} }
\def\AN{{\it Astron. Nachr.} }
\def\IJMP{{\it Int. J. of Mod. Phys.} }
\def\JGR{{\it Journ. of Geophys. Res.}}
\def\JHEP{{\it Journ. of High En. Phys.} }
\def\JPhG{{\it Journ. of Physics} {\bf G} }
\def\MNRAS{{\it Month. Not. Roy. Astr. Soc.} }
\def\Nature{{\it Nature} }
\def\NewAR{{\it New Astron. Rev.} }
\def\PASP{{\it Publ. Astron. Soc. Pac.} }
\def\PhFl{{\it Phys. of Fluids} }
\def\PLB{{\it Phys. Lett.}{\bf B} }
\def\PR{{\it Phys. Rev.} }
\def\PRD{{\it Phys. Rev.} {\bf D} }
\def\PRL{{\it Phys. Rev. Letters} }
\def\RMP{{\it Rev. Mod. Phys.} }
\def\Science{{\it Science} }
\def\ZfA{{\it Zeitschr. f{\"u}r Astrophys.} }
\def\ZfN{{\it Zeitschr. f{\"u}r Naturforsch.} }
\def\etal{{\it et al.}}

\hyphenation{mono-chro-matic sour-ces Wein-berg
chang-es Strah-lung dis-tri-bu-tion com-po-si-tion elec-tro-mag-ne-tic
ex-tra-galactic ap-prox-i-ma-tion nu-cle-o-syn-the-sis re-spec-tive-ly
su-per-nova su-per-novae su-per-nova-shocks con-vec-tive down-wards
es-ti-ma-ted frag-ments grav-i-ta-tion-al-ly el-e-ments me-di-um
ob-ser-va-tions tur-bul-ence sec-ond-ary in-ter-action
in-ter-stellar spall-ation ar-gu-ment de-pen-dence sig-nif-i-cant-ly
in-flu-enc-ed par-ti-cle sim-plic-i-ty nu-cle-ar smash-es iso-topes
in-ject-ed in-di-vid-u-al nor-mal-iza-tion lon-ger con-stant
sta-tion-ary sta-tion-ar-i-ty spec-trum pro-por-tion-al cos-mic
re-turn ob-ser-va-tion-al es-ti-mate switch-over grav-i-ta-tion-al
super-galactic com-po-nent com-po-nents prob-a-bly cos-mo-log-ical-ly
Kron-berg Berk-huij-sen}
\def\simle{\lower 2pt \hbox {$\buildrel < \over {\scriptstyle \sim }$}}
\def\simge{\lower 2pt \hbox {$\buildrel > \over {\scriptstyle \sim }$}}
\def\intunits{{\rm s}^{-1}\,{\rm sr}^{-1} {\rm cm}^{-2}}


\title{The origin of cosmic rays: Explosions of massive stars with 
magnetic winds and their supernova mechanism
}

\author{Peter L.~Biermann\altaffilmark{1,2,3,4}}
\affil{MPI for Radioastronomy, Bonn, Germany}
\email{plbiermann@mpifr.mpg.de}
\author{Julia K.~Becker}
\affil{Ruhr-Universit\"at Bochum, Fakult\"at f\"ur Physik \&
  Astronomie, Theoretische Physik IV, D-44780 Bochum, Germany}
\author{Jens Dreyer}
\affil{Ruhr-Universit\"at Bochum, Fakult\"at f\"ur Physik \&
  Astronomie, Theoretische Physik IV, D-44780 Bochum, Germany}
\author{Athina Meli}
\affil{ECAP, Physik. Inst. Univ. Erlangen-N{\"u}rnberg, Germany}
\author{Eun-Suk Seo}
\affil{Dept. of Physics, Univ. of Maryland, College Park, MD, USA}
\and
\author{Todor Stanev}
\affil{Bartol Research Inst., Univ. of Delaware, Newark, DE, USA}
 
\altaffiltext{1}{FZ Karlsruhe, and Phys. Dept., Univ. Karlsruhe, Germany}
\altaffiltext{2}{Dept. of Phys. \& Astr., Univ. of Alabama, Tuscaloosa, AL, USA}
\altaffiltext{3}{Dept. of Phys., Univ. of Alabama at Huntsville, AL, USA}
\altaffiltext{4}{Dept. of Phys. \& Astron., Univ. of Bonn, Germany }

\begin{abstract}
One prediction of particle acceleration in the supernova
remnants in the magnetic wind of exploding Wolf Rayet and
Red Super Giant stars is that the final spectrum is a
composition of a spectrum $E^{-7/3}$ and a polar cap component
of $E^{-2}$ at the source.
This polar cap component contributes to the total energy content with only a
few percent, but dominates the spectrum at higher energy.
The sum of both components gives spectra which curve upwards.
The upturn was predicted to occur always at the same rigidity.
An additional component of cosmic rays from acceleration by
supernovae exploding into the Inter-Stellar Medium (ISM) adds another
component for Hydrogen and for Helium.
After transport the predicted spectra $J(E)$ 
for the wind-SN cosmic rays
are $E^{-8/3}$ and $E^{-7/3}$; the sum leads to an 
upturn from the steeper spectrum.
An upturn has
now been seen. Here, we test the
observations against the predictions, and show that the observed
properties are consistent with the predictions. Hydrogen can be shown to also have a noticeable
wind-SN-component. The observation of the upturn in the heavy
element spectra being compatible with the same rigidity for all heavy elements
supports the magneto-rotational mechanism for these
supernovae. This interpretation predicts
 the observed upturn to continue to curve upwards and
 approach the $E^{-7/3}$ spectrum. If confirmed, this
would strengthen the case that supernovae of very massive
stars with magnetic winds are important sources of Galactic cosmic
rays.
\end{abstract}

\section{Introduction}

Since 2008 there has been increasing evidence for an extra component
of cosmic ray electrons and positrons, with a plateau of cosmic ray
electrons showing an $E^{-3}$ spectrum, and the CR-positron/electron
ratio approaching $E^{+1/3}$ (ATIC: Chang et al. 2008; Pamela:
Adriani et al. 2009). These results confirm a quantitative model
originally proposed in 1993 (Stanev et al. 1993), in which the
supernova shock racing through a magnetic wind (Biermann 1951, Parker 1958)
gives rise to a source spectrum of $E^{-2}$ from the polar cap of the
star, where the magnetic field is radial, and $E^{-7/3}$ from most of
$4 \pi$, where the magnetic field is nearly circular (Biermann 1993).
Allowing for losses (Kardashev 1962) the polar cap component spectrum
becomes $E^{-3}$ for observers of cosmic ray electrons.
Since acceleration is slower for a radial magnetic field (Jokipii 1987;
Meli \& Quenby 2003a, b; Ellison \& Double 2004; Meli \& Biermann 2006)
than for a tangential magnetic field, more secondaries are produced in
the polar cap region, any resulting secondary spectrum is
shifted down in energy and up in flux thus explaining the observations
(Biermann et al. 2009b).

The prediction of the $E^{-3}$ cosmic ray electron component was
confirmed (HESS: Aharonian et al. 2009) with a spectrum of
$E^{-3.1 \pm 0.1 (stat.) \pm 0.4 (syst.)}$. This approach has also
led to an understanding of the WMAP haze (Finkbeiner 2004;
Hooper et al. 2007) with a predicted spectral behavior of $\nu^{-2/3}$,
$\nu^{-1}$, and $\nu^{-3/2}$ (Biermann et al. 2010). This prediction
is consistent with the detection of the Inverse Compton component
of the WMAP haze, the Fermi haze and its spectrum (Dobler et al. 2010;
Su et al., 2010). These spectral index predictions are derived from
the assumption, that a diffusive regime in a disk connects to a
convective regime in a magnetic wind for cosmic ray transport.
Thus, several independent 
observations support the quantitative model proposed in 1993 (Stanev et 
al. 1993) of massive star explosions. The original cosmic ray spectra 
are consistent will all these new observations.

There are a number of ideas that have suggested an upturn in the
  cosmic ray spectra, such as a nearby source or substructure in the
  accelerating shocks.
Here, we focus on the upturn of the spectra predicted in 1993
(Biermann 1993, Stanev et al. 1993, parallel papers are summarized in 
Biermann 1994).

It had been proposed (see also Prantzos 1984, 1991, 2010;
V{\"o}lk \& Biermann 1988) that the origin of cosmic rays can be
traced to three source components:
a) The cosmic rays originating from supernovae exploding into
the ISM;
b) The cosmic rays from supernovae exploding into red supergiant
(RSG) winds; and c) The cosmic rays from supernovae exploding into
blue supergiant or Wolf Rayet (WR) star winds.
Disregarding the possible binary character of the stars, this is
essentially a mass sequence, and a key property of these stellar
winds is that they are magnetic (Abbott et al. 1984; Barvainis et al.
1987; Drake et al. 1987; Churchwell et al. 1992). As Parker (1958)
showed, the asymptotic magnetic field topology is radial in a polar
cap, and tangential over most of $4 \pi$. Furthermore, observations
of these stars and their interpretation (e.g. Woosley et al. 2002,
Heger et al. 2003) show that the wind eats back down into the star,
and so the abundances in the RSG star winds are enriched in Helium,
and those of WR stars are concentrated in Carbon and Oxygen.
Therefore it can be expected, that in cosmic rays at higher energy much
of the Helium can be traced to these winds of RSG stars,
and possibly all of the elements heavier than Carbon to the WR stars
(Stanev et al. 1993, Biermann 1994).

We use here the concept that interstellar transport (e.g., Rickett 1977)
is governed by a Kolmogorov spectrum (Kolmogorov 1941a, b, c),
as discussed in Biermann (1993), and Biermann et al. (2001).
We test this prediction with new data (CREAM: Seo et al. 2008;
Ahn et al. 2009, 2010). The data are shown to be consistent with
the prediction of 1993, and so we propose further tests.

\section{The upturn observed in cosmic ray spectra}

The general concept developed in 1993 is that, after transport is allowed
for, the spectra are composed of a sum of a $E^{-8/3}$ component,
and a $E^{-7/3}$ polar cap component, which is at a level of a few percent
in integrated power. These components run up to the knee energy, with about $Z \, 10^{15}$ eV, where the polar cap component cuts off completely,
and the main component turns steeper, at about $E^{-3.15}$
(Biermann 1994, Biermann et al. 2005). The spectrum finally cuts off
completely at
about $Z \, 10^{17}$ eV. Here $Z$ is the charge of the cosmic ray
nucleus.

It is not obvious a priori, whether RSG star winds should have the
same properties in terms of their magnetic field as WR star winds:
It is not clear, as to whether the energy fraction of the polar cap
component is the same for both, and also whether the characteristic
knee rigidity ratio is the same. However, for simplicity we will
assume that the population energy fraction and the knee particle
energy are the same for both kinds of stars until the data force us
to accept otherwise.

We first deal with all elements from Carbon (C) through Iron (Fe) and
with the combined data as a function of energy/nucleon. Next we deal
with Helium, and test whether the same numbers can describe its
spectrum. Finally we deal with Hydrogen, which does have a major
ISM-SN contribution. In our 1993 prediction Hydrogen had a cross-over
with Helium, when the spectra are plotted as a function of energy per
particle, which is confirmed now. We also test whether Helium or
any other elements also require an ISM component to understand its spectrum.

\subsection{Carbon through Iron}
The Ahn et al. (2010) data are plotted as $E_n^{5/2} \, J(E_n)$, where $E_n$
is the energy per nucleon. Since the prediction is that $J(E_n)$ is
the sum of one component with $E_n^{-8/3}$ and the polar cap component
with $E_n^{-7/3}$, we note that the product
$E_n^{5/2} \, (E_n^{-8/3} + E_n^{-7/3})$ is a symmetric shallow
near-parabola, with the break at the energy $E_{n, m}$, where the
two contributions have equal flux. Fitting the shape of this function
involves one parameter, that energy per nucleon at the break energy $E_{n, m}$.

First of all, the spectra before the break are determined by CREAM
(Ahn et al. 2009) to be $E^{-2.66 \pm 0.04}$, to be compared with
the 1993 prediction of $E^{-8/3 - 0.02 \pm 0.02}$, with an asymmetric
predicted error distribution. The 2009 determination is reasonably
close to the prediction, as the difference is only 0.007 plus the
corresponding errors. So we stick with $E_n^{-8/3}$. The cosmic ray
electron data suggest in turn, that at source $E^{-2}$ is a good fit
for the polar cap component, and so we also stick with $E_n^{-7/3}$
after transport.

Fitting this to the curves of the elements reveals that the break energy
$E_{n, m}$ is consistent with being the same for all these elements.

The fit curve $E_n^{5/2} \, (E_n^{-8/3} + E_n^{-7/3})$ to the plot
demonstrates, that a break energy of 10 TeV/N is consistent with
the data. The present data are in fact compatible also with 3 or
30 TeV/N to within the errors, but 10 TeV/N is the best common fit.
The 1993 arguments gave 12 TeV/N, derived from a fit to 
overall air shower data, at vertical and slanted zenith angles.

\begin{figure}[htpb]
\centering
\includegraphics[bb=0cm 0cm 15.cm 26cm,viewport=0cm 1.0cm 15cm 
26cm,clip,scale=0.7]{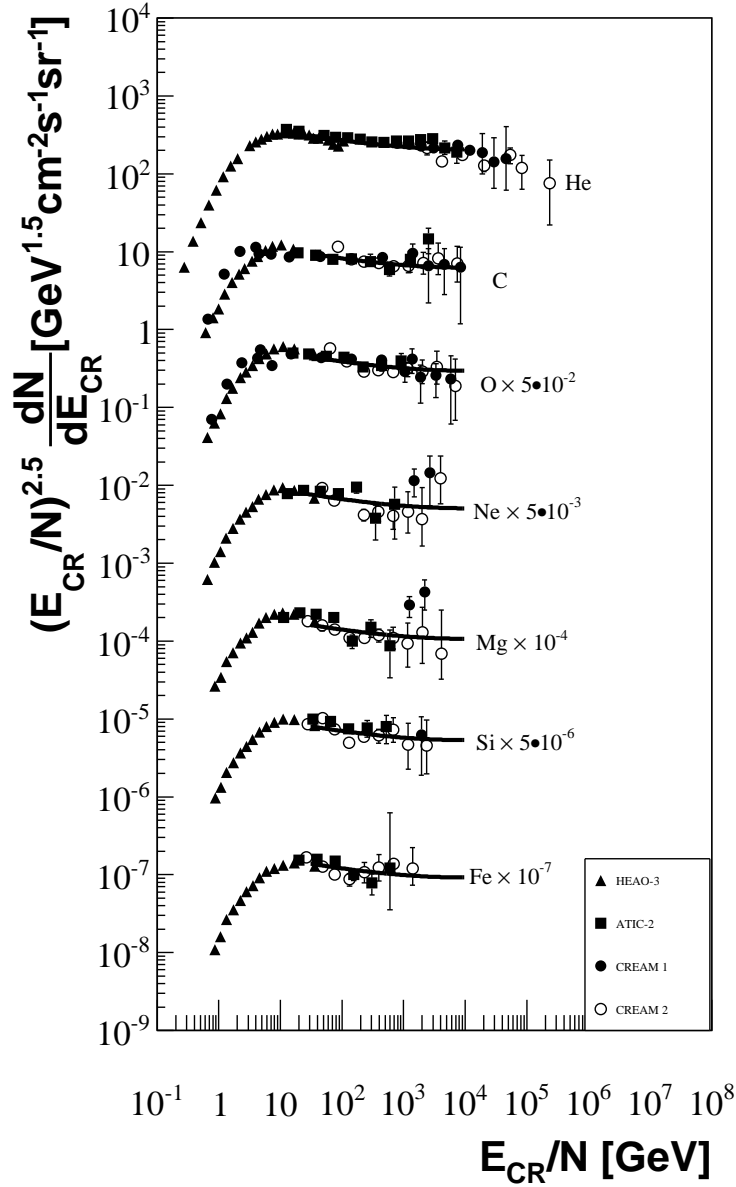}
\caption{Fit for elements heavier than H with the spectral shape from
Stanev et al. (1993). We use for the break energy 10 TeV/N and only fit
the fluxes above 50 GeV/N because of solar modulation and low energy
spallation effects. No ISM-SN CR-component is necessary to fit the spectra.
Lower energy data implies a small ISM-SN CR component for He.}
\label{Allfit}
\end{figure}

We show in Fig.~\ref{Allfit} the fits of all elements heavier than Hydrogen
compared to experimental data. 

\subsection{Helium and Hydrogen}
Measured in flux per energy/particle Helium crosses over Hydrogen near
about 10 TeV. The specific
spectral index of the ISM-contribution has to be determined best by
the Hydrogen component, and that approaches $E^{-2.78 \pm 0.009}$ for
protons at low energy (AMS: Alcaraz et al. 2000).
Since there is a slight flattening due to the wind-component, we need
to adopt a spectrum for the ISM-component, originally predicted to be
$E^{-2.75 \pm 0.04}$, for which we write more generally $E^{-p_{ISM}}$.

\begin{figure}[htpb]
\centering
\includegraphics[bb=0cm 0cm 20cm 13cm,viewport=0cm 0cm 20cm 
13cm,clip,scale=0.8]{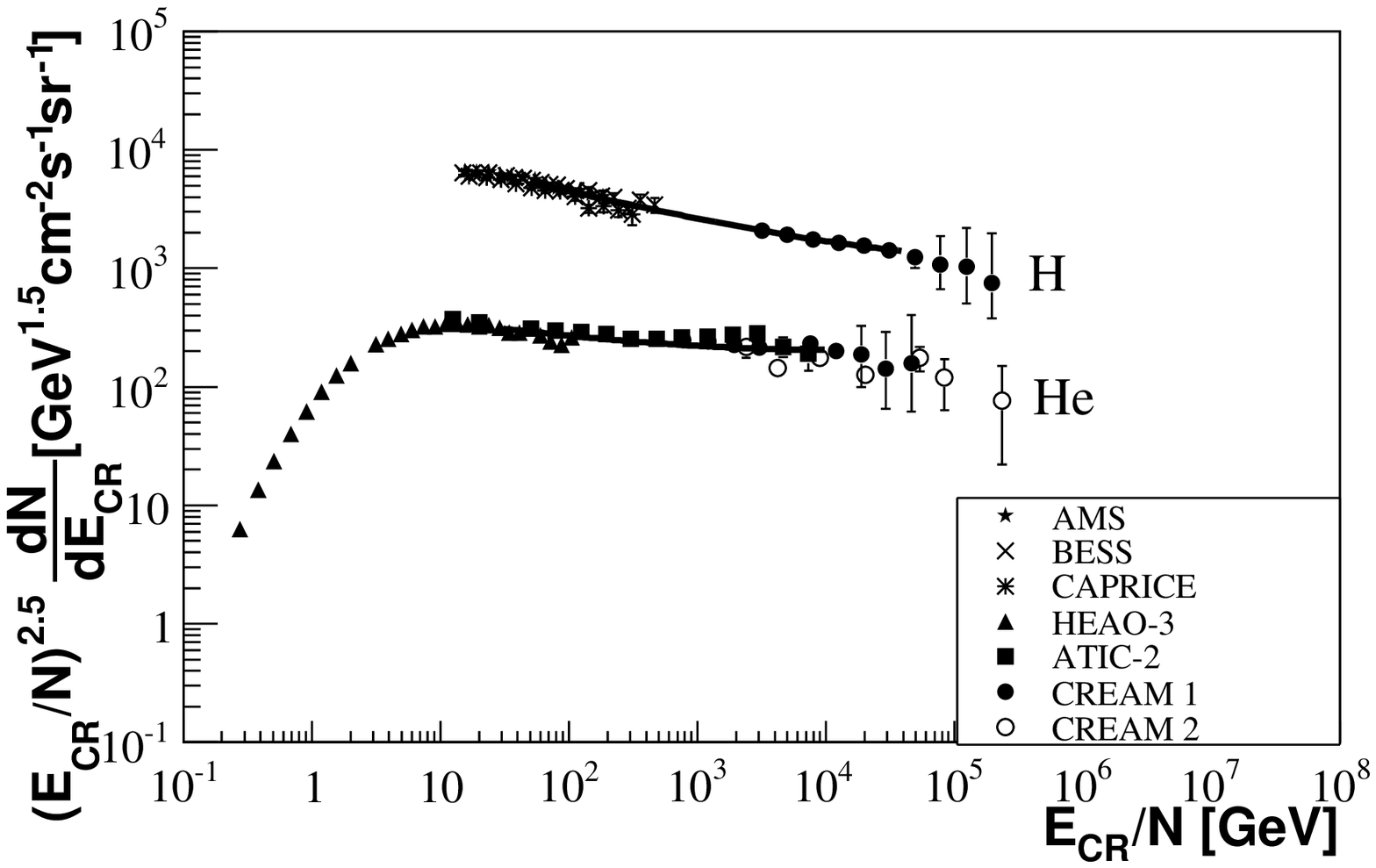}
\caption{Fit to proton and Helium data with spectral shape given in Stanev
et al. (1993). For He we use for the break energy 10 TeV/N, and for
H 20 TeV, since the rigidity E/Z is given. For H the strongest component
is from ISM-SN, a wind-component is required to fit the data, but is
weaker. Replotting these spectra in energy per particle would shift He
up by a factor of 8 and to the right by a factor of 4, at 10 TeV/N,
and so give the cross-over near that energy with Hydrogen.}
\end{figure}
\label{PlotHHe}

Fitting then $E_n^{5/2} \, (E_n^{-8/3} + E_n^{-7/3} + b \, E_n^{-p_{ISM}})$
gives us an estimate of the ISM contribution. We show the fit for H
in Fig~\ref{PlotHHe}, and for comparison repeat the fit for He.

Fitting the data (from AMS: Alcaraz et al. 2000) gives at higher energy a spectral index of $2.66 \, \pm \, 0.02$,
This suggests that
the ISM contribution has $p_{ISM} \, \simeq 2.80$ and then
corresponds to $b_{H} \, \simeq \, 1 $ for Hydrogen at the steeper
spectral index 2.80, and about $b_{H} \, \simeq \, 4 $ for the
slightly flatter index of 2.75. This corresponds to a wind-contribution
of the energetics for CR-Hydrogen of 25 \% for the steeper index,
and about 12 \% for the slightly flatter index. Using
$p_{ISM} \, = \, 2.70$ does not give a satisfactory fit to the
Hydrogen data. The differences illustrate the uncertainties.
For Helium the ISM-CR component is very much smaller, and hard to
define within the errors, using data from independent instruments.

The largest modification with respect to 1993 is the slightly larger contribution
of the wind component to Hydrogen.

\subsection{Differential spectra}
Differential spectra are shown in Ahn et al. (2009). All of those show
near constant ratios, with the exception of Nitrogen: The Nitrogen/Oxygen
ratio shows a weak dependence, consistent with $E^{-5/9}$ suggesting
interaction in an environment of a massive wind-shell, where the spectrum
of magnetic irregularities excited by cosmic ray particles themselves
dominates over other spectra (Biermann 1998; Biermann et al. 2001, 2009b).

The Iron/Oxygen ratio shows a weak increase consistent with weak
differential spallation: Low energy Fe has stronger spallation, so
less iron is left. If this is the true interpretation, then at higher
energy the ratio should approach a constant. The numbers suggest that
up to about half of the low energy iron is lost through spallation.

\subsection{Supernova mechanism}
The rather remarkable observation, that this break energy $E_{n,m}$,
where the $E^{-8/3}$ and the polar cap component $E^{-7/3}$ have
equal fluxes, is well defined, even if not precisely known,
immediately shows that all participating supernovae appear to approach
asymptotically a commonality of correlated magnetic field strength
and explosion energy (Biermann 1993). Stellar evolution of massive
stars (see, e.g., Woosley et al. 2002) suggests, that indeed a
commonality of properties in their asymptotic state might well be
quite plausible. Using the Parker limit
(Parker 1958) for a tangential azimuthal magnetic field this requires
$B_{\phi}(r) \, r \, U_{SN}^{2}$ to be the same for all such
supernovae, and since in this limit $B_{\phi}(r) \, r$ is a constant,
using the base of that regime $r_s$ requires
$B_{\phi , s} \, r_s \, U_{SN}^{2}$ to be the same; here $U_{SN}$ is the
shock speed of the supernova racing through the wind, and
$B_{\phi} \, \sim \, 1/r$ is the azimuthal magnetic field.
Assuming that the mass ejected is similar this connects the explosion
energy directly to the magnetic field strength. The only mechanism
that does that is the magneto-rotational mechanism (Kardashev 1964,
Bisnovatyi-Kogan 1970, Bisnovatyi-Kogan \& Moiseenko 2008, Moiseenko et 
al. 2010).
One could speculate that the neutrinos couple to the exploded
magnetized outflow through instabilities and irregularities, and so
enhance the explosion just as photons can couple better to a wind in
early type stars since the wave speed in a magnetic wind is higher
(Seemann \& Biermann 1997); thus it is conceivable that neutrinos play
a concomitant role in the explosion mechanism
(see, e.g., Woosley et al. 2002).

Since this mechanism employs rotation and magnetic field to their limits,
this mechanism also gives rise to the axial symmetry seen in some 
supernova explosions and required by
Gamma Ray Bursts (GRBs). So it is tempting to speculate that this is in
fact the same mechanism that leads to supernovae and GRBs.
\subsection{Higher cosmic ray energies}
The sum of all these spectra should obey the observed total spectra
of particles determined in other ways. This should be consistent
with the overall spectra all through EeV energies. This test was
used in 1993, and now, with increased H and He wind-contributions,
can be done again. Using the new data increased the Hydrogen and
Helium wind contribution, and so we may have to decrease the break
energy by a small amount to compensate in the total from the knee and 
beyond.

We note that the particles near the knee, here with a notable contribution
also from Hydrogen and Helium, may constitute the seed particles
for the next step up in energy, when a relativistic shock (Gallant \& 
Achterberg 1999) may push the entire spectrum up by a considerable factor. Ultra high energy cosmic rays could be drawn
from a strongly enriched composition (Biermann et al. 2009a,
Gopal-Krishna et al. 2010). With such a concept the results here
suggest that in different parts of the sky, ultra high energy cosmic
rays might be dominated by low mass and very high mass elements
(Abbasi et al. 2009; Abraham et al. 2010).
\section{Conclusions}
Here we demonstrate that the upturn recently observed in cosmic ray
spectra matches the quantitative predictions from 1993.
As now a number of independent observations (CR-positrons, 
CR-electrons, WMAP haze, Fermi haze, the 511 keV line emission from the 
Galactic Center region, and the CR-upturn observed by CREAM) have been 
shown to be consistent with the original quantitative proposal (Stanev 
et al. 1993), the implications originally suggested find support as well.

Since the upturn is consistent with being at the same energy/charge
ratio for all heavy elements, and clearly very nearly the same for
the very large number of participating supernovae, this finding
supports the magneto-rotational mechanism for massive star
explosions (originally proposed by Bisnovatyi-Kogan 1970; see also
Biermann 1993).
Only if energy and magnetic field of the exploding star are strongly 
correlated,
can such a characteristic feature in the observed cosmic ray spectrum
be the same for all such stars, whose explosions contribute to the
cosmic rays at high energy.

A clear prediction is that the upturn should continue to approach
asymptotically a spectrum of $E^{-7/3}$.

As so often, much better data, extending to yet higher energy, will
allow to refine and test the concept at yet higher accuracy.

\section{Acknowledgements}

PLB would like to thank G. Bisnovatyi-Kogan, L. Clavelli, T.K. Gaisser, 
B. Harms, A. Heger, P. Joshi, K.H. Kampert, Gopal-Krishna, N. Langer, S. 
Moiseenko, B. Nath, G. P\u{a}v\u{a}la\c{s}, B. Sadoulet, E. Salpeter, R. 
Sina, V. de Souza, P. Wiita, and many others for discussion of these 
topics, and O. Ofoha for help of the analysis. Support for work with PLB 
has come from the AUGER membership and theory grant 05 CU 5PD 1/2 via 
DESY/BMBF and VIHKOS. JKB and JD acknowledge the support from the 
Research Department of Plasmas with Complex Interactions (Bochum). Support for ESS 
comes from NASA grant NNX09AC14G and for TS comes from DOE grant 
UD-FG02-91ER40626.


\begin{thebibliography}{1}
\expandafter\ifx\csname natexlab\endcsname\relax\def\natexlab#1{#1}\fi
\expandafter\ifx\csname url\endcsname\relax
  \def\url#1{\texttt{#1}}\fi
\expandafter\ifx\csname urlprefix\endcsname\relax\def\urlprefix{URL }\fi
\bibitem[Abbasi et al. (2010)]{a} Abbasi, R. U., et al. (HiRes-Coll.), \PRL {\bf 104}, id. 
161101 (2010); arXiv:0910.4184; with a correction in \PRL {\bf 104}, id. 
199902 (2010).
\bibitem[Abbott et al. (1984)]{b} Abbott, D. C., Bieging, J. H., \& Churchwell, E., \ApJ {\bf 
280}, 671 - 678 (1984)
\bibitem[Abraham et al. (2010)]{c} Abraham, J., et al. (Auger-Coll.), \PRL {\bf 104} id. 091101 
(2010); arXiv:1002.0699
\bibitem[Adriani et al. (2009)]{d} Adriani, O., et al. (Pamela Coll.), \Nature {\bf 458}, 607 - 
609 (2009); arXiv 0810.4995
\bibitem[Aharonian et al. (2009)]{e} Aharonian, F., et al., (H.E.S.S.-Coll.), \AA {\bf 508}, 561 - 
564 (2009); arXiv:0905.0105
\bibitem[Ahn et al. (2009)]{f} Ahn, H.S. et al. (CREAM-Coll.), \ApJ {\bf 707}, 593 - 603 
(2009); arXiv:0911.1889
\bibitem[Ahn et al. (2010)]{g} Ahn, H.S. et al. (CREAM-Coll.), \ApJL {\bf 714}, L89 - L93 
(2010); arXiv:1004.1123
\bibitem[Alcaraz et al. (2000)]{h} Alcaraz, J., (AMS-Coll.) et al., \PLB {\bf 494}, 193 - 202 (2000)
\bibitem[Barvainis et al. (1987)]{i} Barvainis, R., McIntosh, G., Predmore, C. R., \Nature
{\bf 329}, 613 - 615 (1987)
\bibitem[Biermann (1951)]{j} Biermann, L.F., \ZfA {\bf 29}, 274 (1951)
\bibitem[Biermann (1993)]{k} Biermann, P.L., \AA {\bf 271}, 649 (1993) - paper CR-I; 
astro-ph/9301008
\bibitem[Biermann(1994)]{l} Biermann, P.L., 23rd ICRC, in Proc. ``Invited, Rapporteur and 
Highlight papers"; Eds. D. A. Leahy et al., World Scientific, Singapore, 
p. 45 (1994)
\bibitem[Biermann (1998)]{m} Biermann, P.L., Proc. Nuclear Astrophysics meeting at 
Hirschegg, in Proc., GSI, Darmstadt, Germany, p. 211 - 222 (1998)
\bibitem[Biermann et al. (2001)]{n} Biermann, P.L., et al., \AA {\bf 369}, 269 - 277 (2001)
\bibitem[Biermann et al. (2005)]{o} Biermann, P.L., Bisnovatyi-Kogan, G., \& Moiseenko, S., Proc. 
Brazil Nov 2004 meeting, {\it Magnetic fields in the Universe: From 
Laboratory and Stars to Primordial Structures}, Ed. E. d. Pino Gouveia 
\etal, AIP Proc. {\bf 784}, 385 - 395 (2005)
\bibitem[Biermann et al. (2009)]{p} Biermann, P.L., et al., in "High- Energy Gamma-rays and 
Neutrinos from Extra-Galactic Sources", Heidelberg Jan 2009, {\it Int. 
J. of Mod. Phys. D} {\bf 18}, 1577-1581 (2009a); arXiv:0904.1507
\bibitem[Biermann et al. (2009)]{q} Biermann, P. L., Becker, J. K., Meli, A., Rhode, W., Seo, 
E.-S., \& Stanev, T., \PRL {\bf 103}, 061101 (2009b); arXiv:0903.4048
\bibitem[Biermann et al. (1970)]{r} Biermann, P.L., Becker, J.K., Caceres, G., Meli, A., Seo, 
E.-S., \& Stanev, T., \ApJL {\bf 710}, L53 - L57 (2010); arXiv:0910.1197
\bibitem[Bisnovatyi-Kogan (1970)]{s} Bisnovatyi-Kogan, G. S., {\it Astron. Zh.} {\bf 47}, 813 (1970)
\bibitem[Bisnovatyi-Kogan \& Moiseenko (2008)]{t} Bisnovatyi-Kogan, G. S., Moiseenko, S. G., {\it Chinese J. of 
Astron. \& Astroph. Suppl.} {\bf 8}, 330 - 340 (2008)
\bibitem[Chang et al. (2008)]{u} Chang, J., (ATIC-Coll.) et al. \Nature {\bf 456}, 362 (2008)
\bibitem[Churchwell et al. (1992)]{v} Churchwell, E., Bieging, J. H., et al., \ApJ {\bf 393}, 329 - 
340 (1992)
\bibitem[Dobler et al. (2010)]{w} Dobler, G., Finkbeiner, D. P., Cholis, I., Slatyer, T. R., 
Weiner, N., \ApJ {\bf 717}, 825 - 842 (2010); arXiv:0910.4583
\bibitem[Drake et al. (1987)]{x} Drake, St. A., et al. \ApJ {\bf 322}, 902 - 908 (1987)
\bibitem[Ellison \& Double (2004)]{y} Ellison, D.C., \& Double, G.P., \ApP {\bf 22}, 323 (2004)
\bibitem[Finkenbeiner (2004)]{z} Finkbeiner, D.P., \ApJ {\bf 614}, 186 - 193 (2004); 
arXiv:0311547
\bibitem[Gallant \& Achterberg (1999)]{aa} Gallant, Y.-A., \& Achterberg, A., \MNRAS {\bf 305}, L6 - L10 
(1999)
\bibitem[Gopal-Krishna et al. (2010)]{zz} Gopal-Krishna, Biermann, P.L., de Souza, V., Wiita, P.J.,
 \ApJL {\bf 720}, L155 - L158 (2010); arXiv:1006.5022
\bibitem[Heger et al. (2003)]{bb} Heger, A., et al., \ApJ {\bf 591}, 288 - 300 (2003)
\bibitem[Hooper et al. (2007)]{cc} Hooper, D., et al., \PRD {\bf 76}, ms. 083012 (2007)
\bibitem[Jokipii (1987)]{dd} Jokipii, J. R., \ApJ {\bf 313}, 842 - 846 (1987)
\bibitem[Kardashev (1962)]{ee} Kardashev, N.S., {\it Astron. Zh.} {\bf 39}, 393 (1962); 
translated in {\it Sov. Astron.} {\bf 6}, 317 (1962)
\bibitem[Kardashev (1964)]{ff} Kardashev, N. S., {\it Astron. Zh.} {\bf 41}, 807 (1964); 
translated in {\it Sov. Astron.} {\bf 8}, 643 (1965)
\bibitem[Kolmogorov (1941)]{gg} Kolmogorov, A.N., {\it Dokl. Akad. Nauk SSSR} {\bf 30}, 299 - 
303 (1941a); {\bf 31}, 538 - 541 (1941b); {\bf 32}, 19 - 21 (1941c)
\bibitem[Meli (2003)]{hh} Meli, A. \& Quenby, J., \ApP {\bf 19}, 637 (2003a); {\bf 19}, 
649 (2003b)
\bibitem[Meli (2006)]{ii} Meli, A., \& Biermann, P. L., \AA {\bf 454},
p. 687 - 694 (2006); astro-ph/0602308
\bibitem[Moiseenko etl a. (2010)]{jj} Moiseenko, S. G., Bisnovatyi-Kogan, G. S., Ardeljan, N., in 
{\it Astronomy and Beyond: Astrophysics, Cosmology, Radio Astronomy, 
High Energy Physics and Astrobiology}. AIP Conf. Proc. {\bf 1206}, 282 - 
292 (2010)
\bibitem[Parker (1958)]{kk} Parker, E.N., \ApJ {\bf 128} 664 (1958)
\bibitem[Prantzos (1984)]{ll} Prantzos, N., {\it Adv. Space Res.} {\bf 4}, 109 - 114 (1984)
\bibitem[Prantzos (1991)]{mm} Prantzos, N., in Proc. {\it Wolf-Rayet Stars and 
Interrelations with Other Massive Stars in Galaxies:}, Eds. Karel A. van 
der Hucht and Bambang Hidayat, Kluwer, Dordrecht. I.A.U. Symp.{\bf 143}, 
550 (1991)
\bibitem[Prantzos (2010)]{nn} Prantzos, N, , in Proc. IAU Symposium 268 {\it Light Elements 
in the Universe}, Eds. C. Charbonnel, M. Tosi, F. Primas and C. 
Chiappini, I.A.U. Symp. {\bf 268}, in press (2010); arXiv:1003.2317
\bibitem[Rickett (1977)]{oo} Rickett, B. J., \ARAA {\bf 15}, 479 - 504 (1977)
\bibitem[Seemann (1997)]{pp} Seemann, H., \& Biermann, P.L., \AA {\bf 327}, 273 (1997); 
astro-ph/9706117
\bibitem[Seo (2008)]{qq} Seo, E.S., et al. (CREAM-coll.), {\it Adv. Space Res.} {\bf 
42}, 1656 - 1663 (2008)
\bibitem[Stanev et al. (1993)]{rr} Stanev, T., Biermann, P.L. \& Gaisser, T.K., \AA {\bf 274}, 
902 (1993) - paper CR-IV; astro-ph/9303006
\bibitem[Su et al. (2010)]{ss} Su, M., Slatyer, T. R., Finkbeiner, D. P., submitted (2010); 
arXiv:1005.5480
\bibitem[V{\"o}lk \& Biermann (1988)]{tt} V{\"o}lk, H.J., Biermann, P.L., \ApJL {\bf 333}, L65 - L68 (1988)
\bibitem[Woosley et al. (2002)]{uu} Woosley, S. E., Heger, A., Weaver, T. A., \RMP {\bf 74}, 1015 - 1071 (2002)
\end{thebibliography}
\end{document}